\documentclass[11pt]{article}
\usepackage{amssymb}
\usepackage{colortbl}
\usepackage{amsfonts,amsmath, longtable}

        \def\theequation{\thesection.\arabic{equation}}

\newcommand{\tr}{{\rm tr}}
\newcommand{\ti}[1]{\tilde{#1}}

\newcommand{\mO}{{\mathcal O}}

\newcommand{\la}{\lambda}

\newcommand{\be}{\beta}

\newcommand{\Mat}{ {\rm Mat}_N }
\newcommand{\MatM}{ {\rm Mat}(M,\mathbb C) }

\newcommand{\mC}{\mathbb C}
\newcommand{\mZ}{\mathbb Z}

\newcommand{\mS}{\mathcal S}

\newenvironment{proof}{\par\noindent{\bf Proof.}}{\hfill$\scriptstyle\blacksquare$}

\def\beq{\begin{equation}}
\def\eq{\end{equation}}
\def\p{\partial}

\newcommand{\mats}[4]{\left(\begin{array}{cc}{#1}&{#2}\\ {#3}&{#4}
\end{array}\right)}

\def\res{\mathop{\hbox{Res}}\limits}

\begin{document}

\setcounter{page}{1}

\begin{center}

\

\vspace{-5mm}

{\LARGE{\bf  Interrelations between dualities }}

\vspace{3mm}

{\LARGE{\bf in classical integrable systems and classical-classical }}

\vspace{3mm}

{\LARGE{\bf version  of quantum-classical duality}}



 \vspace{18mm}

 {\Large {R. Potapov}} $\,^{\diamond\,\bullet}$
\qquad\quad\quad
 {\Large {A. Zotov}}
 $\,^{\diamond\,\bullet}$

  \vspace{10mm}


$\diamond$ --
 {\em Steklov Mathematical Institute of Russian
Academy of Sciences,\\ Gubkina str. 8, 119991, Moscow, Russia}


$\bullet$ --
{\em Institute for Theoretical and Mathematical Physics,\\ Lomonosov Moscow State University, Moscow, 119991, Russia}


 {\small\rm {e-mails: trikotash@ya.ru, zotov@mi-ras.ru}}

\end{center}

\vspace{0mm}

\begin{abstract}
We describe the Ruijsenaars' action-angle duality in classical many-body integrable systems through
the spectral duality transformation relating the classical spin chains and Gaudin models.
For this purpose, the Lax matrices of many-body systems are represented in the multi-pole (Gaudin-like)
form by introducing a fictitious spectral parameter. This form of Lax matrices is also interpreted
as classical-classical version of quantum-classical duality.
\end{abstract}

\bigskip
%

{\small{ \tableofcontents }}



\section{Introduction}\label{sec1}
\setcounter{equation}{0}

In this paper we consider three types of dualities in integrable systems and describe
certain interrelations between them.
The first one is the Ruijsenaars duality \cite{p-q} known also as action-angle or p-q duality.
It relates integrable many-body systems of the Calogero-Moser-Sutherland \cite{Calogero2} and
Ruijsenaars-Schneider \cite{RS} types by interchanging the action variables and positions of particles.
An easiest way to describe this relation is to use a matrix equation (arising as some moment map equation) for
the Lax matrix of one or another model. For example, the matrix equation
\beq\label{w01}
\displaystyle{
 [A,B]=\mu_0\,,\qquad A,B,\mu_0\in\Mat
  }
\eq
(with some special choice of $\mu_0$) can be solved with respect $A$ by diagonalizing $B$ or vice-versa.
These two possibilities provide Lax matrices of dual rational Calogero-Moser models. Similar
matrix equations can be written for the Lax matrices of the trigonometric Calogero-Sutherland model and the rational Ruijsenaars-Schneider model (namely, $A-B^{-1}AB=\mu_0$), and for a pair of dual
trigonometric Ruijsenaars-Schneider models (namely, $AB^{-1}A^{-1}B=e^{\mu_0}$).

The second duality under consideration is the spectral duality relating the classical Gaudin models and the
classical spin chains. A model of this kind is described by either Lax matrix or a (classical) monodromy matrix $L(z)\in\Mat$ depending on the spectral parameter $z\in\mC$. Its spectral curve
\beq\label{w02}
\displaystyle{
\det\limits_{N\times N}(\lambda-L(z))=0
  }
\eq
is a 1-dimensional complex curve in $\mC^2$ parameterized by $(\lambda,z)$. Coefficients behind $\lambda^k$ in the
left hand side of (\ref{w02}) are conserved quantities. The dual model
is given by some ${\tilde L}(\lambda)\in\MatM$ of possibly different size $M\times M$ depending on the spectral parameter $\lambda$. The spectral duality assumes that the spectral curve of the dual model
\beq\label{w03}
\displaystyle{
\det\limits_{M\times M}(z-{\tilde L}(\lambda))=0
  }
\eq
coincides with (\ref{w02}). First time it was observed for the periodic $N$-body Toda model, which has two Lax representations \cite{FT}. The first representation is $\Mat$-valued and another one is $2\times 2$ representation in the form of
integrable chain. So that, in this case $M=2$, and the dual model coincides with the original one.
Later the spectral duality was formulated for the rational Gaudin models \cite{AHH} and isomonodromic
problems for the rational Schlesinger systems and Painlev\'e equations \cite{Harnad2}, see also \cite{GR}.
Then these results were extended to the spectral duality between the (trigonometric) XXZ Gaudin model and the
(rational) XXX generalized spin chain \cite{MMZZ,MMZZR1}. It was also explained in \cite{MMZZ} that the
duality relation arises naturally in a special limit from the AGT correspondence between some
supersymmetric gauge theory and conformal field theory. Finally, the duality
between a pair of XXZ spin chains was described in \cite{MMZZR2}.

The third duality is the quantum-classical duality between the classical many-body systems of Calogero-Ruijsenaars family and the quantum spin chains or Gaudin models \cite{GZZ,BLZZ}.
Very close phenomena were also previously discussed in \cite{Kasman,MTV2,NRS}. As an example consider
the classical $N$-body rational Calogero-Moser model and the twisted ${\rm gl}_2$ XXX
Gaudin model with the twist matrix $v={\rm diag}(v_1,v_2)$. That is on the classical side we deal with
the system of particles with positions $q_i$ and generalized velocities ${\dot q}_i$, $i=1,...,N$. On the quantum side
we consider the rational Gaudin model on Riemann sphere with $N$ punctures $z_i$, and the eigenvalues of quantum Gaudin Hamiltonians ${\hat H}_i$
computed by means of the algebraic Bethe ansatz method are $h_i$, $i=1,...,N$.
Then the identification between parameters of two models is given by relations
\beq \label{w04}
\begin{array}{l}
\displaystyle{
\nu=\hbar\,,
}
\\
\displaystyle{
q_i = z_i\,,
}
\\
\displaystyle{
\dot{q}_i = h_i/\hbar\,,
}
\end{array}
\eq
where $\nu$ is the Calogero-Moser coupling constant and $\hbar$ is the Planck constat in the quantum Gaudin model.
Also, the action variables of the Calogero-Moser are equal to
\beq \label{w05}
\displaystyle{
\mathrm{Spec} \Big(L^{\hbox{\tiny{RS}}}\Big) = \Big(\underbrace{v_1,...,v_1}_{N-N_1},\underbrace{v_2,...,v_2}_{N_1} \Big),
}
\eq
where $N_1$ is (the number of Bethe roots) the number of the excited spins in the Gaudin model. In this respect
the quantum-classical duality is a relation between the spectrum of quantum spin chains and
certain Lagrangian submanifolds in the phase space of classical integrable systems. This relation holds true
between the classical Ruijsenaars-Schneider model and quantum XXX spin chain. Also, it is naturally
extended to trigonometric many-body systems, and on the quantum side one should consider the XXZ models \cite{BLZZ}.
In principle, one can use it to define the spectrum of quantum spin chain without usage of the Bethe ansatz.
Indeed, from (\ref{w04}) it follows that $h_i=\nu\dot{q}_i$, that is one should find the velocities
of particles for the positions fixed as $q_i=z_i$ and the action variables fixed as in (\ref{w05}).
However, it appears that the duality is extended to one-to-one correspondence when on the quantum side
supersymmetric versions of the chain are taken into account (i.e. those with the group ${\rm GL}_{1|1}$ and
${\rm GL}_{0|2}$ in the above example) \cite{TZZ}.

The topic of dualities is too wide to make even a brief review of all relations. Here we mainly focus
on the descriptions of the Ruijsenaars duality in classical many-body systems and the
spectral duality in the classical spin chains and Gaudin models. See \cite{GVZ,MM,Koroteev} and references therein for more details and applications.

{\bf Purpose of the paper.} The Ruijsenaars duality and the spectral duality looks similar at quantum level, where
both are represented in the form of some bispectral problem. At the same time these dualities deal with different type models.
The first one relates many-body integrable systems, while the second one relates Gaudin models and spin chains.
Moreover, the classical descriptions are absolutely different. The Ruijsenaars duality uses matrix equations for the Lax representations without spectral parameter, while the spectral duality is based on the Lax representations with spectral parameters and the spectral curves (\ref{w02})-(\ref{w03}).

Our first goal is to show that the formulation of spectral dualities based on coincidence
of the spectral curves is applicable to the Ruijsenaars duality as well. For this purpose
we will introduce a fictitious spectral parameter by applying the following gauge transformation to
the Lax matrix (without spectral parameter) of a many-body system:
\beq \label{w06}
\displaystyle{
L \rightarrow L(z)=(z-Q)L(z-Q)^{-1}\,,\qquad Q={\rm diag}(q_1,...,q_N)\,.
}
\eq
The spectral parameter is indeed fictitious since it obviously can be removed by the inverse of the gauge transformation (\ref{w06}),
and the spectral curve (\ref{w02}) is independent of $z$ because it is gauge invariant.
However, using (\ref{w06}) together with some more transformations, the matrix $L(z)$
can be brought to the form of the Lax matrix
(or monodromy matrix) for some special Gaudin model (or spin chain), that is to the matrix-valued function
of $z$ with simple poles at positions of particles. Then by applying the spectral duality transformation
one gets the dual model, which will be shown to be gauge equivalent to the p-q dual many-body system (dual to the original one described by $L$). It should be mentioned that the similarity between Ruijsenaars and spectral dualities is expected from several viewpoints. Firstly, both dualities take the form of certain bispectral problems at quantum level. Secondly, both of these
dualities are just manifestations of the same ${\rm SL}(2,\mZ)$ symmetry of the Ding-Iohara-Miki algebra \cite{Awata}.

Our second goal is to describe an analogue of relations (\ref{w04}) when the spin chain (or Gaudin model) become classical.
The quantum-classical duality relates the classical systems from the first duality and
the quantum models from the second one. An origin of the quantum-classical duality is the quasi-classical limit of the Matsuo-Cherednik projection \cite{FeV,ZZ}
describing wave functions of quantum integrable many-body problems in terms of solutions to (differential or difference) Knizhnik-Zamolodchikov equations. As it was explained by N. Reshetikhin \cite{Resh} the differential Knizhnik-Zamolodchikov equations can be viewed as quantization of the Schlesinger system. At the classical
level the Schlesinger system is a non-autonomous version of the classical Gaudin model. It describes the
isomonodromy problem on ${\mathbb CP}^1/\{z_1,...,z_N\}$, and the marked points $z_1,...,z_N$ (which are simple poles of the Lax connection) play the role of time variables.  Using (\ref{w06}) we will represent
the Lax matrix of the rational Calogero-Moser model in the form of the Schlesinger connection.
In this way we obtain the classical analogue of the identification
of parameters (\ref{w04}), which provides the classical-classical counterpart of the quantum-classical
duality.

The paper is organized as follows. In two next Sections we review the Ruijsenaars and the spectral duality.
In Section 4 we suggest description of the Ruijsenaars duality through the spectral duality transformation.
Finally, in Section 5 we describe the classical-classical version of the quantum-classical
duality. The Tables (\ref{w201}), (\ref{w34}), (\ref{w50}) and (\ref{w501}) presented in this paper are borrowed from \cite{GVZ}.


\section{Recalling Ruijsenaars duality in classical many-body systems}\label{sec2}
\setcounter{equation}{0}

In this Section we recall the duality relations between integrable many-body systems of Calogero-Moser
and Ruijsenaars-Schneider types:
\beq\label{w201}
\begin{array}{|c|c|c|}
\hline
p\diagdown q & rational & trigonometric
\\
\hline
rat. & \hbox{rational CM} & \hbox{trig. CM}
\\
 & \hbox{(self-dual)}\circlearrowleft & \!\nearrow\ \quad\quad\quad\quad\quad \hfill\,
\\
\hline
trig. &\ \hfill\,\quad\quad\quad\quad\quad \swarrow\! & \hbox{trig. RS}
\\
 & \hbox{rational RS} & \hbox{(self-dual)}\circlearrowleft
\\
\hline
\end{array}
\eq
On the horizontal line we put the type of dependence on positions of particles, and
the dependence on momenta is on the vertical
line.

We follow the original algebraic  approach \cite{p-q} based on the properties of Lax pairs and the corresponding matrix equations (see also \cite{AMMZ}).
Although we use some ideas of the group-theoretical approach we do not go into details of this
description, which can be found in
\cite{Arut,Feher2}.

Equations of motion for $N$-body models under consideration have the Lax representation (free of spectral parameter)
in $N\times N$ matrices:
\beq \label{w202}
\displaystyle{
{\dot L}=[L,M]\,,
}
\eq
where $L,M\in\Mat$ are certain matrix valued functions on the phase space of the corresponding model.
Consider the eigenvalue problem
\beq \label{w203}
\displaystyle{
L\Psi=\Psi\Lambda\,,\qquad \Psi\in\Mat\,,\quad \Lambda={\rm diag}(\lambda_1,...,\lambda_N)\in\Mat\,,
}
\eq
where $\Psi$ is a matrix of eigenvectors and $\Lambda$ is a diagonal matrix of eigenvalues.
We assume that $L$ depends on the particles momenta $p_1,...,p_N$, positions of particles $q_1,...,q_N$
and a coupling constant.
Notice that the conservation laws $\tr L^k$, $k\in\mZ_+$ coming from the Lax equation (\ref{w202}) are symmetric functions (the Newton power sums) of eigenvalues
\beq \label{w204}
\displaystyle{
\tr L^k=\tr\Lambda^k=\sum\limits_{i=1}^N\la_i^k\,,
}
\eq
that is $\la_i$ are action variables.

 Suppose the dual model is described by the Lax matrix $\ti L$ depending on ${\ti p}_1,...,{\ti p}_1$,
 ${\ti q}_1,...,{\ti q}_N$. By definition, the duality is an anticanonical map
\beq \label{w205}
\displaystyle{
\omega=-\ti\omega\,,\qquad \omega=\sum\limits_{i=1}^Ndp_i\wedge dq_i\,,
\quad \ti\omega=\sum\limits_{i=1}^Nd{\ti p}_i\wedge d{\ti q}_i
}
\eq
between phase spaces of a pair of models, which identifies the action variables of one model
with positions of particles from another one:
\beq \label{w206}
\begin{array}{c}
\displaystyle{
\lambda_i={\ti q}_i\,,\quad i=1,...,N,
}
\\ \ \\
\displaystyle{
{\ti\lambda}_i={q}_i\,,\quad i=1,...,N.
}
\end{array}
\eq
Introduce the diagonal matrices
\beq \label{w207}
\begin{array}{c}
\displaystyle{
Q={\rm diag}(q_1,...,q_N)\in\Mat\,,\qquad
{\ti Q}={\rm diag}({\ti q}_1,...,{\ti q}_N)\in\Mat\,.
}

\end{array}
\eq
The description of duality, which we use in this paper is based on the eigenvalues problems
\beq \label{w208}
\begin{array}{c}
\displaystyle{
L\Psi=\Psi f({\ti Q})
}
\end{array}
\eq
and
\beq \label{w209}
\begin{array}{c}
\displaystyle{
{\ti L}\Psi^{-1}=\Psi^{-1} {\ti f}(Q)\,,
}
\end{array}
\eq
where $f$ and $\ti f$ are some functions, which will be fixed below.

Below we recall main examples of the duality by demonstrating receipt (\ref{w208})-(\ref{w209})
at the level of matrix equations. At the same time we do not discuss a proof the anticanonicity
(\ref{w205}). This important part of the duality can be found in \cite{p-q,Arut}.

\subsection{Self-duality of the rational Calogero-Moser model}\label{sec21}
The Lax matrix of the rational Calogero-Moser model is
 \beq\label{w210}
 \displaystyle{
L^{\hbox{\tiny{CM}}}_{ij}
=\delta_{ij} p_i+\nu
(1-\delta_{ij})\frac{1}{q_i-q_j}\,, \qquad i,j=1...N\,,
 }
  \eq
where $\nu\in\mC$ is the coupling constant. It is easy to see that this matrix satisfies
equation
 \beq\label{w211}
  \displaystyle{
[Q,L^{\hbox{\tiny{CM}}}]=\nu\bar{\mathcal O}\,,\qquad \bar\mO_{ij}=1-\delta_{ij}\,.
}
 \eq
Equivalently, the matrix $\bar\mO\in\Mat$ is written as
 \beq\label{w212}
  \displaystyle{
\bar\mO=e^T\otimes e-1_N\,,
}
 \eq
where $e$ is the row-vector $e=(1,...,1)$, $e^T$ is the corresponding column-vector, and $1_N$ is the identity matrix.

From viewpoint of (\ref{w208})-(\ref{w209}) in this case we choose $f(x)=x$ and ${\ti f}(x)=x$ in (\ref{w208})-(\ref{w209}), that is $\Psi^{-1}L^{\hbox{\tiny{CM}}}\Psi=\ti Q$
and $\Psi^{-1}Q\Psi={\ti L}^{\hbox{\tiny{CM}}}$. Then, by conjugating both sides of (\ref{w211}) with $\Psi^{-1}$ we get
 \beq\label{w213}
  \displaystyle{
[\ti Q,{\ti L}^{\hbox{\tiny{CM}}}]=-\nu({\ti e}^T\otimes {\breve e}-1_N)\,,
\qquad
{\ti e}^T=\Psi^{-1}e\,,\quad {\breve e}=e\Psi\,.
}
 \eq
The l.h.s. of (\ref{w213}) is an off-diagonal matrix (i.e. $[\ti Q,\ti L]_{ii}=0$).
Therefore, ${\ti e}_i=1/{\breve e}_i$ and solution to (\ref{w213}) yields
 \beq\label{w214}
 \displaystyle{
{\ti L}^{\hbox{\tiny{CM}}}_{ij}
=\delta_{ij} {\ti p}_i-\nu
(1-\delta_{ij})\frac{{\breve e}_j}{{\breve e}_i}\frac{1}{{\ti q}_i-{\ti q}_j}\,, \qquad i,j=1...N\,,
 }
  \eq
Finally, by performing additional gauge transformation ${\ti L}^{\hbox{\tiny{CM}}}\rightarrow D {\ti L}^{\hbox{\tiny{CM}}} D^{-1}$, $D={\rm diag}(d_1,...,d_N)$ and $d_i={\breve e}_i=\sum\limits_k \Psi_{ki}$,
one gets\footnote{The additional gauge transformation is a fixation of freedom in definition of $\Psi$, which
can be multiplied by any diagonal matrix from the right, i.e. $\ti L$ in (\ref{w209}) is defined up to conjugation with a diagonal matrix.}
 \beq\label{w215}
 \displaystyle{
{\ti L}^{\hbox{\tiny{CM}}}_{ij}
=\delta_{ij} {\ti p}_i-\nu
(1-\delta_{ij})\frac{1}{{\ti q}_i-{\ti q}_j}\,, \qquad i,j=1...N\,.
 }
  \eq
This is again the Lax matrix of the rational Calogero-Moser model with the coupling constant $-\nu$.
It can be shown (see \cite{p-q,Arut}) that the anticanonicity condition (\ref{w205}) indeed holds true.
The described phenomenon is called the self-duality of the classical rational Calogero-Moser model.


\subsection{Trigonometric Calogero-Moser-Sutherland
and rational Ruijsenaars-Schneider models}\label{sec22}

The Lax matrix of the Calogero-Moser-Sutherland model has the following form:
 \beq\label{w222}
 \displaystyle{
{L}^{\hbox{\tiny{CMS}}}_{ij}=\delta_{ij} { p}_i + \nu
(1-\delta_{ij})\frac{1}{1-e^{{ q}_i-{ q}_j}}\,, \qquad i,j=1...N\,.
 }
  \eq
It satisfies the matrix equation
 \beq\label{w2191}
 \displaystyle{
L-e^QLe^{-Q}=\nu\bar\mO
 }
  \eq
with the matrix $\bar\mO$ (\ref{w212}).

This time we choose the functions $f$ and $\ti f$ in (\ref{w208})-(\ref{w209}) to be
$f(x)=x$, $\ti f(x)=e^x$, i.e. ${L}^{\hbox{\tiny{CMS}}}=\Psi {\ti Q}\Psi^{-1}$
and $\Psi \ti L \Psi^{-1}= e^Q$. By conjugating (\ref{w2191}) with $\Psi^{-1}$ one gets
 \beq\label{w219}
 \displaystyle{
{\ti Q}-{\ti L}{\ti Q}{\ti L}^{-1}=\nu({\ti e}^T\otimes {\breve e}-1_N)\,,
\qquad
{\ti e}^T=\Psi^{-1}e\,,\quad {\breve e}=e\Psi\,,
 }
  \eq
or, equivalently
 \beq\label{w217}
 \displaystyle{
\nu {\ti L} +[{\ti Q},{\ti L}]=\nu{\ti e}^T\otimes {\breve e}{\ti L} \,.
 }
  \eq
From the diagonal part of this equation we conclude ${\ti e}_i={\ti L}_{ii}/({\breve e}{\ti L})_i$.
By conjugating both sides of (\ref{w217}) with the diagonal matrix ${\rm diag}({\ti e}_1^{-1},...,{\ti e}_N^{-1})$
we obtain
 \beq\label{w2171}
 \displaystyle{
\nu {\ti L} +[{\ti Q},{\ti L}]=\nu{e}^T\otimes {e}\,{\rm diag}({\ti L}) \,,
 }
  \eq
where ${\rm diag}({\ti L})$ is the diagonal part of ${\ti L}$. Solution of (\ref{w2171})
for the off-diagonal part is
 \beq\label{w2172}
 \displaystyle{
{\ti L}_{ij}=\frac{\nu {\ti L}_{jj}}{{\ti q}_i-{\ti q}_j+\nu}\,.
 }
  \eq
The diagonal part remains undefined since the matrix $\ti L$ in (\ref{w219})-(\ref{w2172})
is defined up to multiplication by a diagonal matrix from the right. In order to fix it one should
use analysis of the underlying Poisson structure coming from $T^*{\rm GL}_N$ \cite{p-q,Arut}.
In this way one finally gets
the Lax matrix of the rational Ruijsenaars-Schneider model:
 \beq\label{w2161}
 \displaystyle{
{\ti L}_{ij}={L}^{\hbox{\tiny{rRS}}}_{ij}=\frac{\nu}{{\ti q}_i-{\ti q}_j+\nu}\,e^{{\ti p}_j}b_j({\ti q})\,,
\quad b_j({\ti q})=\prod\limits_{k\neq j}^N\frac{{\ti q}_j-{\ti q}_k-\nu}{{\ti q}_j-{\ti q}_k}\,.
 }
  \eq

\subsection{Self-duality of the trigonometric Ruijsenaars-Schneider model}\label{sec23}
 The Lax matrix of the trigonometric Ruijsenaars-Schneider model is defined as
 \beq\label{w25}
 \displaystyle{
{L}^{\hbox{\tiny{tRS}}}_{ij}=\frac{(1-e^{-\nu})e^{q_i}}{e^{q_i}-e^{-\nu}e^{q_j}}\,e^{p_j}
\prod\limits_{k\neq j}\frac{e^{-\nu}e^{q_j}-e^{q_k}}{e^{q_j}-e^{q_k}}\,,\quad i,j=1,...,N\,.
 }
  \eq
%
%
%
Consider the matrix equation
 \beq\label{w27}
 \displaystyle{
Le^{-Q}L^{-1}e^Q=Ue^{\nu\bar\mO}U^{-1}\,,\qquad \bar\mO=\mO-1_N
 }
  \eq
where $U\in\Mat$ is some invertable matrix. Due to $\mO^2=N\mO$ it is easy to verify that
 \beq\label{w28}
 \displaystyle{
e^{\nu\bar\mO}=e^{-\nu}1_N+\frac{e^{(N-1)\nu}-e^{-\nu}}{N}\,\mO\,.
 }
  \eq
Therefore,
 \beq\label{w281}
 \displaystyle{
Le^{-Q}L^{-1}e^Q=e^{-\nu}1_N+\be a\otimes d\,,\quad \be=\frac{e^{(N-1)\nu}-e^{-\nu}}{N}\,,
 }
  \eq
where $a=Ue^T$ is a column vector and $d=eU^{-1}$ is a row vector. Then
 \beq\label{w282}
 \displaystyle{
L-e^{-\nu}e^{-Q}L e^Q=\be a\otimes d\,e^{-Q}L e^Q\,.
 }
  \eq
Its diagonal part yields $a_i=\be^{-1}(1-e^{-\nu})L_{ii}/(d e^{-Q}L e^Q)_i$. By conjugating the latter
equation with ${\rm diag}(a_1^{-1},...,a_N^{-1})$ we get
 \beq\label{w283}
 \displaystyle{
L-e^{-\nu}e^{-Q}L e^Q=(1-e^{-\nu})\,e^T\otimes e\,{\rm diag}(L)\,.
 }
  \eq
For the off-diagonal part it provides the answer
 \beq\label{w284}
 \displaystyle{
L_{ij}=\frac{(1-e^{-\nu})e^{q_i}}{e^{q_i}-e^{-\nu}e^{q_j}}\,L_{jj}\,,
 }
  \eq
and similarly to the previous case the diagonal part cannot be determined since $L$ in (\ref{w27})
is defined up to multiplication by a diagonal matrix from the right.
An accurate consideration of the underlying Poisson structure
based on the Poisson reduction from the Heisenberg double \cite{Arut} shows that
the final solution has the form  of the Lax matrix (\ref{w25}) of the trigonometric Ruijsenaars-Schneider model.

For the duality transformation the functions $f$ and $\ti f$ in (\ref{w208})-(\ref{w209}) are chosen to be
$f(x)=e^x$, $\ti f(x)=e^x$, i.e. ${L}^{\hbox{\tiny{tRS}}}=\Psi e^{\ti Q}\Psi^{-1}$
and $\Psi \ti L \Psi^{-1}= e^Q$. By conjugating (\ref{w27}) with $\Psi^{-1}$ one gets
 \beq\label{w285}
 \displaystyle{
e^{\ti Q}{\ti L}^{-1}e^{-\ti Q}{\ti L}=\Psi^{-1}Ue^{\nu\bar\mO} U^{-1}\Psi
 }
  \eq
or by taking inverse
 \beq\label{w286}
 \displaystyle{
{\ti L}^{-1}e^{\ti Q}{\ti L}e^{-\ti Q}=\Psi^{-1}Ue^{-\nu\bar\mO} U^{-1}\Psi\,.
 }
  \eq
Finally, conjugating \eqref{w286} also with $e^{-\tilde{Q}} \tilde{L}$, we obtain
 \beq\label{w287}
 \displaystyle{
{\ti L}e^{-\ti Q}{\ti L}^{-1}e^{\ti Q}=e^{-\tilde{Q}}\tilde{L}\Psi^{-1}Ue^{-\nu\bar\mO} U^{-1}\Psi{\ti L}^{-1}e^{\ti Q} = {\ti U}e^{-\nu\bar\mO} {\ti U}^{-1}\,,
 }
  \eq
where
%
 \beq\label{w2871}
{\ti U} = e^{-\tilde{Q}}\tilde{L}\Psi^{-1}U\,.
  \eq
Since $U$ is some invertible matrix, then ${\ti U}$ is also an arbitrary invertible matrix. That is, we see that for  ${\ti L}$ we have the equation of the same form as (\ref{w27}) but
with sign $\nu$ changed. This is the  self-duality of the trigonometric Ruijsenaars-Schneider model.

\section{Recalling spectral duality between classical spin chains and Gaudin models}\label{sec3}
\setcounter{equation}{0}

Spectral dualities are relations between Gaudin models and spin chains \cite{AHH,MMZZ,MMZZR1,MMZZR2}:
\beq\label{w34}
\begin{array}{|c|c|c|}
\hline
\{L\!\stackrel{\otimes}{,}\!L\}_r {\bf \diagdown} r_{12} & \phantom{\Bigl|}XXX\phantom{\Bigl|} & XXZ
\\
\hline
linear & \hbox{rational Gaudin} & \hbox{trig. Gaudin}
\\
 & \hbox{(self-dual)}\circlearrowleft & \!\nearrow\quad\quad\quad\quad\quad\quad\,\hfill
\\
\hline
quadratic & \hfill\,\quad\quad\quad\quad\quad\quad\quad\swarrow\! & \hbox{XXZ spin chain}
\\
 & \hbox{XXX spin chain} & \hbox{(self-dual)}\circlearrowleft
\\
\hline
\end{array}
\eq
On the vertical line we put the type of the classical $r$-matrix structure. The linear one
is
  \beq\label{w330}
  \begin{array}{c}
    \displaystyle{
 \{L_1(z),L_2(w)\}=[L_1(z)+L_2(w),r_{12}(z-w)]
 }
 \end{array}
 \eq
 and the quadratic is
  \beq\label{w331}
  \begin{array}{c}
    \displaystyle{
 \{L_1(z),L_2(w)\}=[L_1(z)L_2(w),r_{12}(z-w)]\,.
 }
 \end{array}
 \eq
 On the horizontal line the type of $r$-matrix (or $L$-matrix) is given, which is rational (XXX) or trigonometric (XXZ).

As we can see the Table (\ref{w34}) is similar to (\ref{w201}).
In fact, from the viewpoint of underlying moment map equations
we could replace the term ''rational'' with ''Lie algebra'', and the term ''trigonometric'' with
''Lie group'' in (\ref{w201}). Being written in this way the table (\ref{w201}) is very similar
to (\ref{w34}) since the Poisson structure is linear on the Lie (co)algebras and it is quadratic on the
Lie groups.

All the models from the Table (\ref{w34}) are represented by some $\Mat$-valued Lax matrix (or the classical monodromy matrix) with $M$ simple poles in spectral parameter. Suppose we deal with a model described by $L(z)\in\Mat$ of the form
  \beq\label{w341}
  \begin{array}{c}
    \displaystyle{
 L(z)=\Lambda+\sum\limits_{k=1}^M\frac{A^k}{z-z_k}\in{\rm Mat}_N\,,\quad A^k\in{\rm Mat}_N\,.
 }
 \end{array}
 \eq
The spectral curve
  \beq\label{w342}
  \begin{array}{c}
    \displaystyle{
 \det\limits_{N\times N}\Big(\lambda-L(z)\Big)=0
 }
 \end{array}
 \eq
in $\mC^2$ (parameterized by $z$ and $\lambda$)
defines all Hamiltonians as coefficients of the polynomial expression (in $z$ or $1/z$ and $\la$ or $1/\la$) in the l.h.s. of (\ref{w342}).

The dual model has the Lax representation with the Lax matrix of size $M\times M$:
  \beq\label{w343}
  \begin{array}{c}
    \displaystyle{
 {\ti L}(\lambda)=
 \ti\Lambda+\sum\limits_{i=1}^{N'}\frac{{\ti A}^i}{\la-\la_i}\in{\rm Mat}_M\,,\quad {\ti A}^i\in{\rm Mat}_M\,.
 }
 \end{array}
 \eq
Consider its spectral curve
  \beq\label{w344}
  \begin{array}{c}
    \displaystyle{
 \det\limits_{M\times M}\Big(z-{\ti L}(\la)\Big)=0\,.
 }
 \end{array}
 \eq
Two models (\ref{w341}) and (\ref{w343}) are called spectrally dual if their spectral curves
(\ref{w342}) and (\ref{w344}) coincide
  \beq\label{w3441}
  \begin{array}{c}
    \displaystyle{
 \det\limits_{N\times N}\Big(\lambda-L(z)\Big)=0\qquad
 \Leftrightarrow\qquad
 \det\limits_{M\times M}\Big(z-{\ti L}(\la)\Big)=0
 }
 \end{array}
 \eq
under certain identification of variables and parameters. The second requirement is the coincidence
of the Seiberg-Witten differentials, see \cite{MMZZ,MMZZR1}.  The latter means that
the identification of variables and parameters of two models provides the Poisson map between them (the Poisson structures for the models under consideration are generated by (\ref{w330}) or (\ref{w331})).
Similarly to the Ruijsenaars duality we will not discuss this condition here but focus on the
matrix (determinant) relations (\ref{w3441}) only.

Notice that the Lax representations are of different sizes, and the variables $z$ and $\lambda$
are interchanged in the expressions for the spectral curves (\ref{w342}) and (\ref{w344}).
This relation is simply understood for the Toda model, which is known to have $2\times 2$ and $N\times N$
Lax representations \cite{FT}.

In particular examples it often happens that $N'=N$ in (\ref{w343}). For this reason the duality is sometimes called the rank-size duality (or $({\rm gl}_N,{\rm gl}_M)$ duality at the level of Knizhnik-Zamolodchikov equations) assuming that the number of sites and the rank of group in spin chain type models
are interchanged under duality transformation \cite{MTV1}.
In fact, the spectral duality we discuss here is a part of a larger phenomenon arising in different areas of theoretical  and mathematical physics. In particular, the duality is naturally lifted to the level of quantum models, 
the Knizhnik-Zamolodchikov equations and gauge theories. See \cite{MMZZ,GVZ,Koroteev,NRS,MMZ} and references therein.

\subsection{Self-duality of the rational (XXX) Gaudin model}
Main computational trick is based on the following well-known determinant relation.
Consider a matrix $\mats{A}{B}{C}{D}$ of size $(N+M)\times(N+M)$, where $A$ is an invertible $N\times N$ square matrix, $D$ is an invertible $M\times M$ square matrix, $B$ is a rectangular $N\times M$ matrix and
$C$ is a rectangular $M\times N$ matrix. Then
  \beq\label{w3443}
  \begin{array}{c}
    \displaystyle{
 \det\limits_{(N+M)\times (N+M)}\mats{A}{B}{C}{D}=
 }
 \\ \ \\
     \displaystyle{
 =\det\limits_{N\times N}(A)\det\limits_{M\times M}\Big(D-CA^{-1}B\Big)=
 \det\limits_{M\times M}(D)\det\limits_{N\times N}\Big(A-BD^{-1}C\Big)\,.
 }
 \end{array}
 \eq
The rational ${\rm gl}_N$ Gaudin model on ${\mathbb CP}^1/\{z_1,...,z_M\}$ is defined by the Lax matrix of the form:
  \beq\label{w345}
  \begin{array}{c}
    \displaystyle{
 L(z)=\Lambda+\sum\limits_{k=1}^M\frac{S^k}{z-z_k}\in{\rm Mat}_N\,,\quad S^k\in{\rm Mat}_N\,,
 }
 \end{array}
 \eq
where $\Lambda={\rm diag}(\la_1,...,\la_N)\in\Mat$ is the diagonal twist matrix. We consider a special case when
all matrices of dynamical variables $S^a$, $a=1,...,M$ are of rank one:
  \beq\label{w346}
  \begin{array}{c}
    \displaystyle{
 (S^k)_{ij}=\xi^k_i\eta^k_j\,,\quad k=1,...,M\,,\quad i,j=1,...,N\,.
 }
 \end{array}
 \eq
Then the dual Gaudin model is defined by the Lax matrix \cite{AHH}:
  \beq\label{w347}
  \begin{array}{c}
    \displaystyle{
 {\ti L}(\la)=Z+\sum\limits_{i=1}^N\frac{A^i}{\la-\la_i}\in{\rm Mat}_M\,,\quad A^i\in{\rm Mat}_M\,,
 }
 \end{array}
 \eq
where $Z={\rm diag}(z_1,...,z_M)\in{\rm Mat}_M$ is a diagonal $M\times M$ twist matrix and
  \beq\label{w348}
  \begin{array}{c}
    \displaystyle{
 (A^i)_{kl}=\xi^k_i\eta^l_i\,,\quad k,l=1,...,M\,,\quad i=1,...,N\,.
 }
 \end{array}
 \eq
Notice that $\xi$ with elements $(\xi)_{ik}=\xi^k_i$ is a rectangular $N\times M$ matrix, and similarly
$\eta$ with elements $(\eta)_{kj}=\eta^k_j$ is a rectangular $M\times N$ matrix.
The statement (\ref{w3441}) follows from direct application of (\ref{w3443}) with
$A=\Lambda-1_N\lambda$, $B=\xi$, $C=\eta$ and $D=Z-z$. Indeed, the Lax matrices (\ref{w345}) and (\ref{w347})
take the following form:
  \beq\label{w349}
  \begin{array}{c}
    \displaystyle{
 L(z)=\Lambda+\xi(z-Z)^{-1}\eta\in{\rm Mat}_N\,,\qquad {\ti L}(\la)=Z+\eta(\la-\Lambda)^{-1}\xi\in{\rm Mat}_M\,.
 }
 \end{array}
 \eq
Then (\ref{w3443}) yields
  \beq\label{w350}
  \begin{array}{c}
    \displaystyle{
 \frac{\det\limits_{N\times N}\Big(\lambda-L(z)\Big)}{\det\limits_{N\times N}\Big(\lambda-\Lambda\Big)}
 =
 \frac{\det\limits_{M\times M}\Big(z-{\ti L}(\la)\Big)}{\det\limits_{M\times M}\Big(z-Z\Big)}\,.
 }
 \end{array}
 \eq

\subsection{Spectral duality between XXZ Gaudin model and XXX spin chain}
\paragraph{Trigonometric Gaudin model.}

Let us now define the Lax matrix of the trigonometric Gaudin model
  \beq\label{w359}
  \begin{array}{c}
    \displaystyle{
 {L^{\hbox{\tiny{tG}}}}'(z)=\Lambda+\sum\limits_{k=1}^M \tr_2\Big(r_{12}(z/z_k)S_2^k\Big)\,,\quad S^k_2=1_N\otimes S^k\,,
 \quad S^k\in\Mat\,,\ k=1,...,M\,,
 }
 \end{array}
 \eq
based on the $r$-matrix
  \beq\label{w3571}
  \begin{array}{c}
    \displaystyle{
 r_{12}(z)
 =\sum\limits_{i,j=1}^N \Big(\frac{1}{z-1}+\delta_{i>j}\Big) E_{ij}\otimes E_{ji}\,.
 }
 \end{array}
 \eq
More precisely,
  \beq\label{w360}
  \begin{array}{c}
    \displaystyle{
 {L^{\hbox{\tiny{tG}}}}'_{ij}(z)=\delta_{ij}\la_i+\sum\limits_{k=1}^M \delta_{i>j}S_{ij}^k+
 \sum\limits_{k=1}^M\frac{z_k}{z-z_k}S_{ij}^k\,.
 }
 \end{array}
 \eq
An explanation of how  $r$-matrix (\ref{w3571}) appears from the original XXZ classical $r$-matrix is given in the
Appendix A.

\paragraph{Trigonometric Gaudin model in the form of the rational untwisted reduced Gaudin model.}
In what follows we consider (similarly to the rational case) the special Gaudin model (\ref{w360}), where
all dynamical matrices $S^k\in\Mat$ are of rank one:
  \beq\label{w3661}
  \begin{array}{c}
    \displaystyle{
 S^k_{ij}=\xi^k_i\eta^k_j\,,\quad k=1,...,M\,,\quad i,j=1,...,N\,.
 }
 \end{array}
 \eq
The second term in the r.h.s. of (\ref{w360}) is a low-triangular matrix.
It was explained in \cite{MMZZR2} that this second term  can be gauged away by applying
certain (triangular) gauge transformation
 ${L^{\hbox{\tiny{tG}}}}'(z)\rightarrow L^{\hbox{\tiny{tG}}}(z)=g^{-1}{L^{\hbox{\tiny{tG}}}}'(z)g$.
 It is constructed as follows. Consider the matrices $S^k$ (\ref{w3661}). Define
  \beq\label{w3610}
  \begin{array}{c}
    \displaystyle{
 \mS=\sum\limits_{k=1}^M S^k\in\Mat\,.
 }
 \end{array}
 \eq
 and the matrix $g\in\Mat$
 with elements
  \beq\label{w3611}
  \begin{array}{c}
    \displaystyle{
 g_{ij}=\delta_{ij}+\frac{\delta_{i>j}}{\la_j-\la_i}
 \sum\limits_{k_i,...,k_{j+1}=1}^M\xi_i^{k_i}
 \Big( \delta_{k_ik_{i-1}}+\frac{ \eta^{k_i}_{i-1}\xi^{k_{i-1}}_{i-1} }{\la_j-\la_{i-1}} \Big)
 \dots
  \Big( \delta_{k_{j+2}k_{j+1}}+\frac{ \eta^{k_{j+2}}_{j+1}\xi^{k_{j+1}}_{j+1} }{\la_j-\la_{j+1}} \Big)
  \eta_j^{k_{j+1}}\,.
 }
 \end{array}
 \eq
 By introducing $M$-dimensional columns $\eta_i=(\eta_i^1,...,\eta_i^M)^T$ and
 $M$-dimensional rows $\xi_i=(\xi_i^1,...,\xi_i^M)$ we have
  \beq\label{w3612}
  \begin{array}{c}
    \displaystyle{
 g_{ij}=\delta_{ij}+\frac{\delta_{i>j}}{\la_j-\la_i}
\xi_i
 \Big( 1_M+\frac{ \eta_{i-1}\otimes\xi_{i-1} }{\la_j-\la_{i-1}} \Big)
 \dots
  \Big( 1_M+\frac{ \eta_{j+1}\otimes\xi_{j+1} }{\la_j-\la_{j+1}} \Big)
  \eta_j\,.
 }
 \end{array}
 \eq
It satisfies the following property. Let ${\bar \mS}$ be a matrix with elements ${\bar \mS}=\delta_{i>j}\mS_{ij}$.
Then
  \beq\label{w362}
  \begin{array}{c}
    \displaystyle{
  g^{-1}(\Lambda+{\bar \mS})g=\Lambda\,.
 }
 \end{array}
 \eq
For readers convenience it is proved in the Appendix B.
In (\ref{w360}) the second term is ${\bar S}^1+...+{\bar S}^M=\bar\mS$. Thus, by applying this gauge
transformation
the second term vanishes.
 Then we come to
  \beq\label{w363}
  \begin{array}{c}
    \displaystyle{
 L^{\hbox{\tiny{tG}}}(z)=\Lambda+
 \sum\limits_{k=1}^M\frac{z_k {\breve S}^k}{z-z_k}\,,\qquad {\breve S}^k=g^{-1}S^k g \in{\rm Mat}_N\,.
 }
 \end{array}
 \eq
Let us mention that in the original model (\ref{w360}) the Poisson brackets for $S^k_{ij}$
are given by the linear Poisson-Lie structure. The Poisson structure for the
 gauged transformed matrices ${\breve S}^k$
is more complicated. The Lax matrix (\ref{w363}) can be obtained
as a result of reduction of the rational Gaudin model on ${\mathbb CP}^1/\{0,z_1,...,z_M,\infty\}$
of the form (\ref{w345}) with $\Lambda=0$. The reduction is by the coadjoint action of ${\rm GL}_N$ Lie group.
One can fix the gauge as $A^\infty=\Lambda$ and solve the moment map equation as $A^0=-\Lambda-\sum_{k=1}^M A^k$.
See details in \cite{MMZZR1}.

In the case (\ref{w3661}) for the set of matrices ${\breve S}^k$ we have
  \beq\label{w3662}
  \begin{array}{c}
    \displaystyle{
 {\breve S}^k_{ij}={\breve\xi}^k_i{\breve\eta}^k_j\,,\quad k=1,...,M\,,\quad i,j=1,...,N\,,
 }
 \end{array}
 \eq
where ${\breve\xi}=g^{-1}\xi$ and ${\breve\eta}=\eta g$. Equivalently,
  \beq\label{w3663}
  \begin{array}{c}
    \displaystyle{
 {\breve S}^k={\breve\xi}^k\otimes{\breve\eta}^k\,,\quad k=1,...,M\,,
 }
 \end{array}
 \eq
where ${\breve\xi}^k$ are $N$-dimensional columns and ${\breve\eta}^k$ are $N$-dimensional rows.

\paragraph{XXX spin chain.} We deal with the generalized twisted
 ${\rm GL}_M$ spin chain on $N$ sites. It is defined through the quadratic $r$-matrix
  structure (\ref{w331}) with
  \beq\label{w364}
  \begin{array}{c}
    \displaystyle{
 r^{\hbox{\tiny{xxx}}}_{12}(\la)=\frac{1}{\la}\,P_{12}\,,\quad
 P_{12}=\sum\limits_{k,l=1}^N e_{kl}\otimes e_{lk}\in{\rm Mat}_M^{\otimes 2}\,,
 }
 \end{array}
 \eq
where $e_{ij}$ is the standard basis in ${\rm Mat}_M$. Using the Lax matrices
  \beq\label{w365}
  \begin{array}{c}
    \displaystyle{
 L^{i}(\la)=1_M+\frac{1}{\la}\,B^i\in{\rm Mat}_M\,,
 }
 \end{array}
 \eq
where $B^i$ are matrices of dynamical variables at $i$-th site, the monodromy matrix takes the form
  \beq\label{w366}
  \begin{array}{c}
    \displaystyle{
 T(\la)=VL^N(\la-\la_N)...L^1(\la-\la_1)=V\Big(1_M+\sum\limits_{i=1}^N\frac{{\ti S}^i}{\lambda-\lambda_i}\Big)
=V+\sum\limits_{i=1}^N\frac{V{\ti S}^i}{\lambda-\lambda_i}\,.
 }
 \end{array}
 \eq
where $V={\rm diag}(v_1,...,v_M)$ is a diagonal twist matrix
and the matrices ${\ti S}^i\in{\rm Mat}_M$ are obtained by expanding the
product of $L$-matrices into the sum, that is $V{\ti S}^i=\res\limits_{\la=\la_i}T(\la)$.
Below we consider the case ${\rm rank}(B^i)=1$. Then ${\rm rank}({\ti S}^i)=1$ since
the expression $\res\limits_{\la=\la_i}T(\la)$ contains $B^i=\res\limits_{\la=\la_i}L^i(\la-\la_i)$ as a factor.

\paragraph{Duality.} Let us show the coincidence of the spectral curves of $N$-site ${\rm GL}_M$ XXX chain
and the trigonometric XXZ ${\rm gl}_N$ Gaudin model with $M$ marked points\footnote{From viewpoint of the rational reduced Gaudin model the number of marked points is $M+2$ including $0$ and $\infty$, see \cite{MMZZR1}.}.
Following \cite{MMZZ,MMZZR1} for the duality transformation we use the relation
  \beq\label{w367}
  \begin{array}{c}
    \displaystyle{
\det(1_N-D)=\exp(\tr(\log(1_N-D)))
 }
 \end{array}
 \eq
and
  \beq\label{w3671}
  \begin{array}{c}
    \displaystyle{
\tr(\log(1_N-D))=-\sum\limits_{k=1}^\infty\frac{1}{k}\,\tr(D^k)
 }
 \end{array}
 \eq
for a matrix $D\in{\rm Mat}_N$. Consider expression for the spectral curve of the trigonometric Gaudin model (\ref{w363}): $\det(\lambda-L^{\hbox{\tiny{tG}}}(z))$. Recall that the Lax matrix (\ref{w363}) is gauge equivalent to the (\ref{w360}). Therefore,
  \beq\label{w368}
  \begin{array}{c}
    \displaystyle{
\det\limits_{N\times N}\Big(\lambda-{L^{\hbox{\tiny{tG}}}}'(z)\Big)
=\det\limits_{N\times N}\Big(\lambda-L^{\hbox{\tiny{tG}}}(z)\Big)\,.
 }
 \end{array}
 \eq
Next,
  \beq\label{w369}
  \begin{array}{c}
    \displaystyle{
\det\limits_{N\times N}\Big(\lambda-L^{\hbox{\tiny{tG}}}(z)\Big)=
\det\limits_{N\times N}\Big(\lambda-\Lambda\Big)
\det\limits_{N\times N}\Big(1_N-(\lambda-\Lambda)^{-1}\sum\limits_{k=1}^M\frac{z_k {\breve S}^k}{z-z_k}\Big)\,.
 }
 \end{array}
 \eq
By expanding the exponent in the r.h.s. of (\ref{w367}) we see that $\det(1_N-D)$ takes the form of a sum
with terms of the form $c_{n_1,...,n_k}\tr(D^{n_1})...\tr(D^{n_k})$. It is important that
the coefficients $c_{n_1,...,n_k}$ are independent of the size of the matrix $N$. Then our strategy is to represent
$\tr_{N\times N}(D^{n_k})$ in the form $\tr_{M\times M}({\ti D}^{n_k})$ for some $M\times M$ matrix $\ti D$.
With this treatment we will have
  \beq\label{w3691}
  \begin{array}{c}
    \displaystyle{
\det\limits_{N\times N}\Big(1_N-D\Big)=\det\limits_{M\times M}\Big(1_M-{\ti D}\Big)\,.
 }
 \end{array}
 \eq
Let
  \beq\label{w3692}
  \begin{array}{c}
    \displaystyle{
D=(\lambda-\Lambda)^{-1}\sum\limits_{k=1}^M\frac{z_k {\breve S}^k}{z-z_k}\in{\rm Mat}_N
 }
 \end{array}
 \eq
 and $Z={\rm diag}(z_1,...,z_M)\in{\rm Mat}_M$.
We have
  \beq\label{w370}
  \begin{array}{c}
    \displaystyle{
\tr_{N\times N}(D^n)=\sum\limits_{i_1,...,i_n=1}^N D_{i_1i_2}...D_{i_ni_1}=
}
\\
    \displaystyle{
=\sum\limits_{i_1,...,i_n=1}^N\sum\limits_{k_1,...,k_n=1}^M
\frac{1}{\la-\la_{i_1}}\frac{z_{k_1}{\breve\xi}^{k_1}_{i_1}{\breve\eta}^{k_1}_{i_2}}{z-z_{k_1}}\dots
\frac{1}{\la-\la_{i_n}}\frac{z_{k_n}{\breve\xi}^{k_n}_{i_n}{\breve\eta}^{k_n}_{i_1}}{z-z_{k_n}}=
 }
\\
    \displaystyle{
\sum\limits_{i_1,...,i_n=1}^N\sum\limits_{k_1,...,k_n=1}^M
\frac{1}{\la-\la_{i_1}}\frac{z_{k_n}}{z-z_{k_n}}{\breve\eta}^{k_n}_{i_1}{\breve\xi}^{k_1}_{i_1}
\frac{1}{\la-\la_{i_2}}\frac{z_{k_1}}{z-z_{k_1}}{\breve\eta}^{k_1}_{i_2}{\breve\xi}^{k_2}_{i_2}
\dots
\frac{1}{\la-\la_{i_n}}\frac{z_{k_{n-1}}}{z-z_{k_{n-1}}}{\breve\eta}^{k_{n-1}}_{i_n}{\breve\xi}^{k_n}_{i_n}=
 }
\\
    \displaystyle{
=\sum\limits_{i_1,...,i_n=1}^N\tr_{M\times M}
\Big(   \frac{1}{\la-\la_{i_1}}(z-Z)^{-1}Z({\breve\eta}_{i_1}\otimes{\breve\xi}_{i_1})
\dots \frac{1}{\la-\la_{i_n}}(z-Z)^{-1}Z({\breve\eta}_{i_n}\otimes{\breve\xi}_{i_n})
 \Big)=
  }
\\
    \displaystyle{
    =\tr_{M\times M}\Big( \sum\limits_{i=1}^N\frac{1}{\la-\la_{i}}(z-Z)^{-1}Z({\breve\eta}_{i}\otimes{\breve\xi}_{i})\Big)\,,
    }
 \end{array}
 \eq
where ${\breve\eta}_i$, $i=1,...,N$ is a set of $M$-dimensional columns and
${\breve\xi}_i$, $i=1,...,N$ is a set of $M$-dimensional rows.
Therefore,
  \beq\label{w371}
  \begin{array}{c}
    \displaystyle{
\ti D=\sum\limits_{i=1}^N\frac{1}{\la-\la_{i}}(z-Z)^{-1}Z({\breve\eta}_{i}\otimes{\breve\xi}_{i})\,.
 }
 \end{array}
 \eq
Plugging $D$ (\ref{w3692}) and $\ti D$ (\ref{w371}) into relation (\ref{w3691}) we obtain
  \beq\label{w372}
  \begin{array}{c}
    \displaystyle{
\det\limits_{N\times N}\Big(\lambda-L^{\hbox{\tiny{tG}}}(z)\Big)=
\frac{\det\limits_{N\times N}(\la-\Lambda)}{\det\limits_{M\times M}(z-Z)}
\det\limits_{M\times M}
\Big[z-Z\Big(1_M+\sum\limits_{i=1}^N\frac{{\breve\eta}_{i}\otimes{\breve\xi}_{i}}{\la-\la_{i}}\Big)\Big]\,.
 }
 \end{array}
 \eq
By comparing it with the expression for the spectral curve of the spin chain $\det_{M\times M}(z-T(\la))$
(\ref{w366}) we obtain the identification of parameters: $V=Z$ and ${\ti S}^i={\breve\eta}_{i}\otimes{\breve\xi}_{i}$.

\subsection{Self-duality of the trigonometric (XXZ) spin chain}
The classical generalized ${\rm GL}_N$ XXZ spin chain on $M$ sites is defined by the monodromy matrix
  \beq\label{w380}
  \begin{array}{c}
    \displaystyle{
 T(z)=VL^M(z/z_M)...L^1(z/z_1)\in{\rm Mat}_N\,,
 }
 \end{array}
 \eq
where $V={\rm diag}(v_1,...,v_N)$ is a diagonal twist matrix, $z_1,...,z_M\in\mC$ are inhomogeneous parameters and
the Lax matrices are constructed by means of the classical XXZ $r$-matrix:
  \beq\label{w381}
  \begin{array}{c}
    \displaystyle{
L^i(z/z_i)=1_N+\tr_2(r_{12}(z/z_i)B^i_2)\in{\rm Mat}_N\,,\quad B^i_2=1_N\otimes B^i\,,\quad B^i\in\Mat\,.
 }
 \end{array}
 \eq
Similarly to the trigonometric Gaudin model we use here the $r$-matrix (\ref{w357}), that is
  \beq\label{w382}
  \begin{array}{c}
    \displaystyle{
L^i(z/z_i)=1_N+\sum\limits_{k=1}^M {\bar B}^k+
 \sum\limits_{k=1}^M\frac{z_kB^k}{z-z_k}\,,\qquad {\bar B}^k=\sum\limits_{i,j=1}^N\delta_{i>j}B_{ij}^kE_{ij}\,.
 }
 \end{array}
 \eq
As in the previous cases we assume ${\rm rank}(B^k)=1$, $k=1,...,M$.

By expanding the product of $L$-matrices into the sum the monodromy matrix is represented in the form
  \beq\label{w383}
  \begin{array}{c}
    \displaystyle{
T(z)=V\Big(1_N+\sum\limits_{k=1}^M {\bar U}^k+
 \sum\limits_{k=1}^M\frac{z_kW^k}{z-z_k}\Big)
 \,,\qquad {\bar U}^k=\sum\limits_{i,j=1}^N\delta_{i>j}a_i^kb_j^kE_{ij}\,,
 }
 \end{array}
 \eq
that is $U^k=a^k\otimes b^k$ with some $N$-dimensional columns $a^k$ and $N$-dimensional rows $b^k$.
Notice also that ${\rm rank}(W^k)=1$ due to
  \beq\label{w384}
  \begin{array}{c}
    \displaystyle{
z_kVW_k=\res\limits_{z=z_k}T(z)=VL^M(z_k/z_M)...L^M(z_k/z_{k+1})z_k B^kL^M(z_k/z_{k-1})...L^1(z_k/z_{1})
 }
 \end{array}
 \eq
and ${\rm rank}(B^k)=1$.
By applying the gauge transformation of type (\ref{w3611})-(\ref{w362}) we may cancel out the second term in
(\ref{w383}).
This allows to represent the monodromy matrix in the form
  \beq\label{w385}
  \begin{array}{c}
    \displaystyle{
T(z)=V\Big(1_N+
 \sum\limits_{k=1}^M\frac{z_kS^k}{z-z_k}\Big)\,,\quad {\rm rank}(S^k)=1\,.
 }
 \end{array}
 \eq
In fact, the proof of the duality between XXZ spin chains does not require to perform the gauge transformation, see \cite{MMZZR2}. But here we will make it similarly to the case of trigonometric Gaudin model.
Consider expression for the spectral curve
  \beq\label{w386}
  \begin{array}{c}
    \displaystyle{
\det\limits_{N\times N}\Big(1_N\la-T(z)\Big)=\det\limits_{N\times N}\Big(1_N\la-V\Big)\det\limits_{N\times N}\Big(1_N
 -(1_N\la-V)^{-1}V\sum\limits_{k=1}^M\frac{z_kS^k}{z-z_k}\Big)\,.
 }
 \end{array}
 \eq
Denote $S^k=\xi^k\otimes \eta^k$ with some $N$-dimensional columns $\xi^k$ and $N$-dimensional rows $\eta^k$.
We are going to use formula (\ref{w3691}) with
  \beq\label{w387}
  \begin{array}{c}
    \displaystyle{
D=(1_N\la-V)^{-1}V\sum\limits_{k=1}^M\frac{z_kS^k}{z-z_k}\in\Mat\,.
 }
 \end{array}
 \eq
Let us compute $M\times M$ matrix $\ti D$:
  \beq\label{w388}
  \begin{array}{c}
    \displaystyle{
\tr_{N\times N}D^{m} = \sum\limits_{i_1,...,i_m = 1}^N\sum\limits_{a_{1},...,a_m = 1}^M
\frac{v_{i_{1}}}{\la-v_{i_{1}}}
\frac{z_{a_{1}}\xi_{i_{1}}^{a_{1}}\eta_{i_{2}}^{a_{1}}}{z-z_{a_{1}}}\frac{v_{i_{2}}}{\la-v_{i_{2}}}
\frac{z_{a_{2}}\xi_{i_{2}}^{a_{2}}\eta_{i_{3}}^{a_{2}}}{z-z_{a_{2}}}...\frac{v_{i_{m}}}{\la-v_{i_{m}}}
\frac{z_{a_{m}}\xi_{i_{m}}^{a_{m}}\eta_{i_{1}}^{a_{m}}}{z-z_{a_{m}}} =
}
\\ \ \\
    \displaystyle{
= \sum\limits_{i_1,...,i_m = 1}^N\sum\limits_{a_{1},...,a_m = 1}^M
\frac{z_{a_{m}}}{z-z_{a_{m}}}\frac{v_{i_{1}}\eta_{i_{1}}^{a_{m}}\xi_{i_{1}}^{a_{1}}}{\la-v_{i_{1}}}
\frac{z_{a_{1}}}{z-z_{a_{1}}}\frac{v_{i_{2}}\eta_{i_{2}}^{a_{1}}\xi_{i_{2}}^{a_{2}}}{\la-v_{i_{2}}}...
\frac{z_{a_{m-1}}}{z-z_{a_{m-1}}}\frac{v_{i_{m}}\eta_{i_{m}}^{a_{m-1}}\xi_{i_{m}}^{a_{m}}}{\la-v_{i_{m}}}
=
}
\\ \ \\
    \displaystyle{
 = \tr_{M\times M}\Big((z 1_M-Z)^{-1}Z\eta V(\la 1_N-V)^{-1}\xi\Big)^{m}\,,
 }
 \end{array}
 \eq
where
  \beq\label{w389}
  \begin{array}{c}
    \displaystyle{
    Z={\rm diag}(z_1,...,z_M)\in{\rm Mat}_M
 }
 \end{array}
 \eq
and $\eta$ is a rectangular $M\times N$ matrix, while $\eta$ is a rectangular $N\times M$ matrix.
In this way we obtain
  \beq\label{w390}
  \begin{array}{c}
    \displaystyle{
    {\ti D}=(z-Z)^{-1}Z\eta V(\la-V)^{-1}\xi\in{\rm Mat}_M\,.
 }
 \end{array}
 \eq
Finally, from (\ref{w3691}) we conclude
  \beq\label{w391}
  \begin{array}{c}
    \displaystyle{
    \frac{\det\limits_{N\times N}\Big(\la-T(z)\Big)}{\det\limits_{N\times N}\Big(\la -V\Big)}=
    \frac{\det\limits_{M\times M}\Big(z-{\ti T}(\la)\Big)}{\det\limits_{M\times M}\Big(z -Z\Big)}\,,
 }
 \end{array}
 \eq
where
  \beq\label{w392}
  \begin{array}{c}
    \displaystyle{
    {\ti T}(\la)=Z\Big(1_M+\eta V(\la-V)^{-1}\xi\Big)=
    Z\Big(1_M+\sum\limits_{i=1}^N\frac{v_i{\ti S}^i}{\la-v_i}\Big)\in{\rm Mat}_M\,,
 }
 \end{array}
 \eq
  \beq\label{w393}
  \begin{array}{c}
    \displaystyle{
 {\ti S}^i=\eta_i\otimes\xi_i\in{\rm Mat}_M\quad \hbox{or}\quad
 {\ti S}^i_{kl}=\eta^k_i\xi^l_i\,,\ i=1,...,N\,,\ k,l=1,...,M\,.
 }
 \end{array}
 \eq

\section{Ruijsenaars duality in the form of spectral duality}\label{sec4}
\setcounter{equation}{0}
In this Section we use the strategy described in the Introduction. Namely, we perform the gauge transformation (\ref{w06}), which allows to rewrite the Lax matrices of many-body systems in the ''Gaudin-like'' forms.
Using also the eigenvalue problems (\ref{w208})-(\ref{w209}) we apply the transformation of spectral duality
$L(z)\rightarrow {\tilde L}(\lambda)$ as was explained in (\ref{w345})-(\ref{w350}). The spectrally dual model appears
to be gauge equivalent to the Ruijsenaars dual Lax matrix.
\subsection{Relation between rational models}
%
Consider the self-duality of the rational Calogero-Moser model, see Subsection \ref{sec21}.  Let us
introduce the fictitious spectral parameter $z$ by performing the gauge transformation applied to the
Lax matrix of the Calogero-Moser model (\ref{w210}):
 \beq\label{rs1}
\displaystyle{
	L'(z) = (z-Q)L^{\hbox{\tiny{CM}}}(z-Q)^{-1}
	=L^{\hbox{\tiny{CM}}} + \sum\limits_{a = 1}^{N}\frac{\nu{\bar\mO}^{a}}{z-q_{a}}\,,\quad \bar\mO^a\in\Mat
}
\eq
where
 \beq\label{rs2}
\displaystyle{
\bar\mO^{a}_{ij} = -(1-\delta_{ij})\delta_{aj}\,,\quad
a,i,j=1,...,N
}
\eq
Indeed, due to
  \beq\label{rs201}
  \begin{array}{c}
    \displaystyle{
 L'_{ij}(z)=\delta_{ij}p_i+\nu\frac{1-\delta_{ij}}{q_i-q_j}\frac{z-q_i}{z-q_j}
 =\delta_{ij}p_i+(1-\delta_{ij})\Big(\frac{\nu}{q_i-q_j}-\frac{\nu}{z-q_j}\Big)\,.
 }
 \end{array}
 \eq
we get (\ref{rs1}).
Next, we apply the gauge transformation with the matrix of eigenvectors $\Psi$ (\ref{w208}):
 \beq\label{rs3}
\displaystyle{
L(z) = \Psi^{-1}L'(z)\Psi = \tilde{Q} - \nu\Psi^{-1}\bar\mO(z-Q)^{-1}\Psi\,,
\quad
\bar\mO= \sum\limits_{a = 1}^{N}\bar\mO^a\in\Mat
}
\eq
with the same $\bar\mO$ as in (\ref{w212}). Equivalently,
 \beq\label{rs301}
\displaystyle{
L(z) = \tilde{Q}+ \sum\limits_{a = 1}^{N}\frac{\nu{\Psi^{-1}\bar\mO}^{a}\Psi}{z-q_{a}}
}
\eq
The spectral parameter is fictitious since it appeared from the gauge transformation.
The expression for the characteristic polynomial (''the spectral curve'')  is independent of $z$:
  \beq\label{rs202}
  \begin{array}{c}
    \displaystyle{
\det(\lambda-L(z))=\det(\lambda-L'(z))=\det(\lambda-L)=\prod\limits_{k=1}^N(\lambda-{\ti q}_k)\,.
 }
 \end{array}
 \eq
Turning back to (\ref{rs301}) we mention that
the residues $\res\limits_{z=q_a}L(z)=\nu{\Psi^{-1}\bar\mO}^{a}\Psi$ are rank one matrices.
They are represented in the form $\res\limits_{z=q_a}L_{ij}(z)=\xi^a_i\eta^a_j$ with
 \beq\label{rs4}
\displaystyle{
	\xi^{a}_{i} = -\nu\sum\limits_{m = 1}^{N}\Psi^{-1}_{im}(1-\delta_{ma})\,,\qquad \eta^{a}_{j} = \Psi_{aj}\,.
}
\eq
%
Therefore, the Lax matrix (\ref{rs301}) takes the form (\ref{w345})-(\ref{w346}) with $M=N$, $q_i=z_i$, $i=1,...,N$, $\Lambda=\ti Q$ and $\xi_i^a$, $\eta_b^j$ are defined as in (\ref{rs4}).

The spectrally dual Lax matrix is constructed by means of (\ref{w347})-(\ref{w348}):
 \beq\label{rs5}
\displaystyle{
	\tilde{L}(\lambda) = Q - \nu\Psi(\la-\tilde{Q})^{-1}\Psi^{-1}\bar\mO.
}
\eq
Taking in to account  (\ref{w208})-(\ref{w209})
we have
\beq\label{rs6}
\displaystyle{
\tilde{L}'(\la) = \Psi^{-1} \tilde{L}(\la)\Psi =
 \tilde{L}^{CM} - \nu(\la-\tilde{Q})^{-1}\Psi^{-1}\bar\mO\Psi\,.
}
\eq
As the last step we use the additional diagonal gauge transformation, as explained in Section 2.1.
This yields
\beq\label{rs7}
\displaystyle{
	\tilde{L}'(\la) = \tilde{L}^{CM} - \nu(\la-\tilde{Q})^{-1}\bar\mO\,.
}
\eq
Gauging this expression with $(z-\tilde{Q})^{-1}$, one obtains nothing but Ruijsenaars dual rational Calogero-Moser model
\beq\label{rs8}
\displaystyle{
 (\la-\tilde{Q})^{-1}\tilde{L}'(z)(\la-\tilde{Q})=\tilde{L}^{CM}\,.
}
\eq

\subsection{Relation between rational and trigonometric models}
In the same manner we treat the trigonometric Calogero-Moser-Sutherland model. Choose
 $f(\tilde{x}) = e^{\tilde{x}}$ and $\tilde{f}(x) = x$
  in the eigenvalue problem and consider the Lax matrix \eqref{w222}.
  Similarly to the previous case we have:
 \beq\label{rs9}
\displaystyle{
	L(z) = \Psi^{-1}(z-e^{Q})^{-1}L^{CMS}(z-e^{Q})\Psi =  e^{\tilde{Q}} - \nu \Psi^{-1}e^{Q} (z-e^{Q})^{-1}e^{-Q}\bar\mO e^{Q}\Psi.
}
\eq
By introducing
 \beq\label{rs10}
\displaystyle{
	\xi^{i}_{a} = -\nu\Psi^{-1}_{ia}, \,\,\,\,\,\, \eta^{i}_{a} = (e^{-Q}\bar \mO e^{Q}\Psi)_{ia},\,\,\,\,\,\, \Lambda = e^{\tilde{Q}} ,\,\,\,\,\,\, Z= e^{{Q}},
}
\eq
the expression (\ref{rs9}) takes the form of the Gaudin model \eqref{w363}:
 \beq\label{rs11}
\displaystyle{
	L(z) = \Lambda + \xi Z(z-Z)^{-1}\eta\,.
}
\eq
Next, using (\ref{w372}) we define the spectrally dual Lax matrix. It has the form the monodromy
in the XXX spin chain:
 \beq\label{rs12}
\displaystyle{
	T'(\lambda) =
e^{\tilde{Q}}\Big(1_N-\nu e^{-Q}\bar \mO e^{Q}\Psi(\lambda-\tilde{Q})^{-1}\Psi^{-1}\Big)\,.
}
\eq
Conjugate it one again with the matrix of eigenvectors $\Psi$:
\beq\label{rs13}
\begin{array}{c}
\displaystyle{
	T(\lambda) = \Psi^{-1}T'(\lambda)\Psi =
L^{rRS}\Big(1_N-\nu \Psi^{-1}e^{-Q}\bar \mO e^{Q}\Psi(\lambda-\tilde{Q})^{-1}\Big)=
}
\\ \ \\
\displaystyle{
=L^{rRS}-\nu L^{rRS}\Psi^{-1}e^{-Q}\bar \mO e^{Q}\Psi(\lambda-\tilde{Q})^{-1}\,.
}
\end{array}
\eq
Consider the second term in detail. Due to $L^{rRS}=\Psi^{-1}e^Q\Psi$ we are left with
\beq\label{rs14}
\displaystyle{
-\nu \Psi^{-1}\bar \mO\Psi L^{rRS}(\lambda-\tilde{Q})^{-1}.
}
\eq
Taking into account the moment map equation (\ref{w219}) and keeping in mind the diagonal gauge transformation (as explained in Section 2.2) we obtain for this term
\beq\label{rs15}
\displaystyle{
	-\nu L^{rRS}\Psi^{-1}e^{-Q}\bar \mO e^{Q}\Psi (\lambda-\tilde{Q})^{-1}=  (-\tilde{Q}L^{rRS}+ L^{rRS}\tilde{Q})(\lambda-\tilde{Q})^{-1}.
}
\eq
Finally, plugging (\ref{rs15}) into (\ref{rs13}) yields
\beq\label{rs16}
\displaystyle{
	T(\lambda) = L^{rRS} + [L^{rRS},\tilde{Q}](\lambda- \tilde{Q})^{-1}\,.
}
\eq
The latter expression is gauge equivalent to the rational Ruijsenaars-Schneider model with
the diagonal gauge transformation:
\beq\label{rs17}
\displaystyle{
	T(\lambda) = (\lambda- \tilde{Q})L^{rRS}(\lambda- \tilde{Q})^{-1}.
}
\eq

\subsection{Relation between trigonometric models}
Description of the self duality of trigonometric Ruijsenaars-Schneider model
 is similar to the previous one.
The only difference is in the choice of functions in the eigenvalue problem. Now $f(\tilde{x}) = e^{\tilde{x}}$ and $\tilde{f}(x) = e^{x}$ in (\ref{w285})-(\ref{w286}).
Let us perform the conjugation by the diagonal matrix to turn the trigonometric Ruijsenaars-Schneider model (\ref{w25}) into the spin chain  form:
\beq\label{rs18}
\displaystyle{
	T'(z) = (z- e^{Q})L^{tRS}(z- e^{Q})^{-1} =
  L^{tRS}\Big(1_N +(L^{tRS})^{-1}[L^{tRS},e^{Q}]e^{Q}(z-e^{Q})^{-1}e^{-Q}\Big)\,.
  }
\eq
Then, after additional conjugation with $\Psi$ the matrix (\ref{rs18}) acquires the form of the  monodromy matrix
(\ref{w385}) of XXZ spin chain
\beq\label{rs19}
\displaystyle{
	T(z) = \Psi^{-1}T'(z)\Psi = V\Big(1_N + \xi Z(z-Z)^{-1} \eta\Big)
}
\eq
with the following identification:
\beq\label{rs20}
\displaystyle{
	V = e^{\tilde{Q}},\,\,\,\, \xi = \Psi^{-1}(L^{tRS})^{-1}[L^{tRS},e^{Q}],\,\,\,\,
\eta = e^{-Q}\Psi,\,\,\,\, Z = e^{Q}}\,.
\eq
Using (\ref{w392})-(\ref{w393}) let us write down the spectrally dual Lax matrix (more precisely, the dual monodromy matrix):
\beq\label{rs21}
\displaystyle{
 \tilde{T}(\lambda) = e^{Q}\Big(1_N + e^{-Q}\Psi e^{\tilde{Q}}(\lambda - e^{\tilde{Q}})^{-1}\Psi^{-1}(L^{tRS})^{-1}[L^{tRS},e^{Q}]\Big)\,.
 }
\eq
The last step is again the conjugation with $\Psi$:
 \beq\label{rs22}
\begin{array}{c}
	\displaystyle{
		\tilde{T}'(\lambda) = 	\Psi^{-1}\tilde{T}(\lambda)\Psi = \tilde{L}^{tRS}+e^{\tilde{Q}}(\lambda-e^{\tilde{Q}})^{-1}\Psi^{-1}(L^{tRS})^{-1}[L^{tRS},e^{Q}] \Psi =
	}
	\\ \ \\
	\displaystyle{
		= \tilde{L}^{tRS} + (\lambda-e^{\tilde{Q}})^{-1}[e^{\tilde{Q}},\tilde{L}^{tRS}]\,.
		}\end{array}
\eq
The Lax matrix for the model is obtained by gauge transformation with the diagonal matrix:
\beq\label{rs23}
\displaystyle{
	\tilde{L}^{tRS}=(\lambda-e^{\tilde{Q}})\tilde{T}'(\lambda)(\lambda-e^{\tilde{Q}})^{-1}\,.
}
\eq

\section{Classical-classical counterpart of quantum-classical duality}\label{sec5}
\setcounter{equation}{0}

\subsection{An origin of quantum-classical duality}
The quantum-classical duality describes a relation between quantum spin chains (or quantum Gaudin models) and classical many-body systems \cite{GZZ,BLZZ}
 \beq\label{w50}
 \begin{array}{c}
 \hbox{\underline{Quantum-classical dualities:}}
 \\ \ \\
 \begin{array}{ccc}
   \hbox{class. rat. Calogero-Moser} & \longleftrightarrow & \hbox{quant. rat. Gaudin}
 \\
   \hbox{class. trig. Calogero-Moser} & \longleftrightarrow & \hbox{quant. trig. Gaudin}
  \\
   \hbox{class. rat. Ruijsenaars-Schneider} & \longleftrightarrow & \hbox{quant. XXX chain}
   \\
   \hbox{class. trig. Ruijsenaars-Schneider} & \longleftrightarrow & \hbox{quant. XXZ chain}
 \end{array}
  \end{array}
 \eq
with the identification of parameters as was given in the Introduction (\ref{w04}) and (\ref{w05}).
The relations $\hbar{\dot q}_i=h_i$ (\ref{w04}) follow from the quasiclassical
limit of the Matsuo-Cherednik projection from the Knizhnik-Zamolodchikov (KZ) equations to quantum many-body
systems \cite{MC,FeV}:
 \beq\label{w501}
 \begin{array}{c}
 \hbox{\underline{Matsuo-Cherednik projections:}}
 \\ \ \\
 \begin{array}{ccc}
 \hbox{rational KZ} & \longrightarrow & \hbox{quant. rat. Calogero-Moser}
 \\
  \hbox{trigonometric KZ} & \longrightarrow &  \hbox{quant. trig. Calogero-Moser}
  \\
  \hbox{rational qKZ} & \longrightarrow &  \hbox{quant. rat. Ruijsenaars-Schneider}
   \\
  \hbox{trigonometric qKZ} & \longrightarrow & \hbox{quant. trig. Ruijsenaars-Schneider}
 \end{array}
  \end{array}
 \eq
These projections are ''quantum-quantum'' versions of the dualities (\ref{w50}).

In this Section we deal with the simplest case corresponding to the upper line in (\ref{w50}) and (\ref{w501}).
The KZ equations can be written in the form:
 \beq\label{w55}
 \begin{array}{c}
  \displaystyle{
  \kappa\p_{z_i}\Phi={\hat H}_i\Phi\,,\quad i=1,...,N\,,
   }
 \end{array}
 \eq
 where $\kappa\in\mC$ is a constant and
  ${\hat H}_i$ are commuting quantum Hamiltonians of a certain Gaudin model. Solution $\Phi$
   to the system of KZ equations is an element of some Hilbert space, where ${\hat H}_i$ act.
 Consider the quasi-classical limits of the KZ equations:
 \beq\label{w502}
 \begin{array}{ccc}
  & \hbox{\underline{KZ equations}}
  &
 \\
  \qquad\qquad\qquad\qquad\qquad\qquad\qquad
 \swarrow 3 &  & 1 \searrow
  \qquad\qquad\qquad\qquad\qquad\qquad\qquad
 \\
   \hbox{class. Schlesinger system} &  & \hbox{quant. Gaudin model}
  \\
 \qquad\qquad\qquad\qquad\qquad\qquad\qquad
   \searrow 4 &  & 2 \swarrow
 \qquad\qquad\qquad\qquad\qquad\qquad\qquad
   \\
    & \hbox{class. Gaudin model} &
 \end{array}
 \eq
The arrow 1 here is the standard quasi-classical limit $\kappa\rightarrow 0$ of (\ref{w55})
corresponding to the expansion $\Phi=(\Phi_0+\kappa\Phi_1+...)\exp(S\kappa)$. It was described in \cite{ReshV}.
As a result one obtains the quantum Gaudin model with the eigenvalue problem ${\hat H}_i\Phi_0=h_i\Phi_0$, $i=1,...,N$. At the same time the Hamiltonians ${\hat H}_i$ depend also on the second Planck constant $\hbar$.
The classical limit of the latter one $\hbar\rightarrow 0$ is the arrow 2 on the Table (\ref{w502}). It maps Lie algebra generators (entering ${\hat H}_i$)
to their symbols, and the commutation relations become the Poisson-Lie brackets.

\subsection{Gaudin model and Schlesinger system}
The limit from the KZ equations to the classical Gaudin model can be achieved in a different way.
Consider the arrow 3 on the Table (\ref{w502}). It was shown in \cite{Resh} (see also \cite{LO}) that the KZ equations can be viewed as quantization of the Schlesinger system.
In the limit $\hbar\rightarrow 0$ together with redefinition $\kappa=\hbar{\ti \kappa}\rightarrow 0$ one gets
the set of Schlesinger equations for the symbols of Lie algebra generators (in our simplest case
we deal with ${\rm gl}_N$ Lie algebra):
%
%
 %
 \beq\label{w56}
 \begin{array}{c}
  \displaystyle{
  \ti\kappa\p_{z_i}S^j=-\frac{[S^i,S^j]}{z_i-z_j}\,,\quad i\neq j\,,\quad i,j=1,...,N\,,
   }
   \\
   \displaystyle{
  \ti\kappa\p_{z_i}S^i=\sum\limits_{k:k\neq i}^N\frac{[S^i,S^k]}{z_i-z_k}+[S^i,\Lambda]\,,\quad i=1,...,N\,,
   }
 \end{array}
 \eq
 where $S^i\in\Mat$ are dynamical variables and $\Lambda\in\Mat$ is the diagonal constant twist matrix.
 The Schlesinger systems is non-autonomous mechanics since the time variables $z_i$ enter in the equations
 of motion explicitly. Its autonomous analogue is give by the Gaudin model with the equations of motion
 \beq\label{w57}
 \begin{array}{c}
  \displaystyle{
  \p_{t_i}S^j=-\frac{[S^i,S^j]}{z_i-z_j}\,,\quad i\neq j\,,\quad i,j=1,...,N\,,
   }
   \\
   \displaystyle{
  \p_{t_i}S^i=\sum\limits_{k:k\neq i}^N\frac{[S^i,S^k]}{z_i-z_k}+[S^i,\Lambda]\,,\quad i=1,...,N\,.
   }
 \end{array}
 \eq
 The time variables $t_i$ correspond to Hamiltonian dynamics with the Hamiltonians
 \beq\label{w58}
 \begin{array}{c}
  \displaystyle{
  H_a=\sum\limits_{c:c\neq a}^N\frac{\tr(S^aS^c)}{z_a-z_c}+\tr(S^a\Lambda)\,,\quad a=1,...,N
   }
 \end{array}
 \eq
 and the Poisson brackets
 \beq\label{w581}
 \begin{array}{c}
  \displaystyle{
\{S^a_{ij},S^b_{kl}\}=\delta^{ab}(-S^a_{il}\delta_{kj}+S^a_{kj}\delta_{il}),\quad a,b,i,j,k,l=1,...,N
   }
 \end{array}
 \eq
 on $({\rm gl}_N^*)^{\times N}$.

 The arrow 4 on the Table (\ref{w502}) is the autonomous limit from the Schlesinger system (\ref{w56}) to the Gaudin model. It is a special procedure \cite{LO}
 including the limit $\ti\kappa\rightarrow 0$.

Equations of motion for the Gaudin model (\ref{w57}) are represented in the Lax form
 \beq\label{w59}
 \begin{array}{c}
  \displaystyle{
  \frac{d}{dt_a}L(z)=[L(z),M_a(z)]\,,\quad a=1,...,N
   }
 \end{array}
 \eq
with
 \beq\label{w60}
 \begin{array}{c}
  \displaystyle{
  L(z)=\Lambda+\sum\limits_{c=1}^N\frac{S^c}{z-z_c}\,,\qquad M_a(z)=-\frac{S^a}{z-z_a}\,.
   }
 \end{array}
 \eq
The same Lax pair reproduces the Schlesinger equations (\ref{w56}) through the monodromy preserving equation
 \beq\label{w61}
 \begin{array}{c}
  \displaystyle{
  {\ti\kappa}\p_{z_a}L(z)-\ti\kappa\p_z M_a(z)=[L(z),M_a(z)]\,,\quad a=1,...,N\,.
   }
 \end{array}
 \eq
Equations of motion for Gaudin model (\ref{w57}) and the Schlesinger systems (\ref{w56})
are Hamiltonian. They are generated by Hamiltonians (\ref{w58}) and Poisson brackets (\ref{w581}).
The Hamiltonians (\ref{w58}) appear in the expansion
 \beq\label{w62}
 \begin{array}{c}
  \displaystyle{
\frac12\,\tr(L^2(z))=\sum\limits_{a=1}^N\frac{C_a}{(z-z_a)^2}+
\sum\limits_{a=1}^N\frac{H_a}{z-z_a}+H_0\,,
   }
 \end{array}
 \eq
 where $C_a=\tr((S^a)^2/2$ are Casimir functions of (\ref{w581}) and
 \beq\label{w63}
 \begin{array}{c}
  \displaystyle{
H_0=\frac12\,\tr(\Lambda^2)
   }
 \end{array}
 \eq
is a constant since $\Lambda$ is a constant diagonal twist matrix.

\subsection{Calogero-Moser model in the form of fictitious Schlesinger system}

Recall that previously in (\ref{rs3}) we represented the Lax matrix of the Calogero-Moser model
in the form of Gaudin model. The identification of parameters is as follows:
 \beq\label{w64}
 \begin{array}{c}
  \displaystyle{
z_i=q_i\,,\quad i=1,...,N\,,
   }
 \end{array}
 \eq
 \beq\label{w65}
 \begin{array}{c}
  \displaystyle{
\Lambda={\ti Q}\,,
   }
 \end{array}
 \eq
that is the diagonal elements of the twist matrix $\Lambda$ are action variables.
This identification is similar to (\ref{w04})-(\ref{w05}) from the quantum-classical duality.
More precisely, (\ref{w64}) is similar to the second line of (\ref{w04}) and (\ref{w65}) is similar to (\ref{w05}).
In this subsection we explain how to deduce a classical analogue of the third line of (\ref{w04}).

We mention that (\ref{rs3}) is not a true Gaudin model since the Poisson brackets
between matrix elements of residues $\mO^a$ are not of the form (\ref{w581}). Moreover, the diagonal elements
of the twist matrix are not constants. These are action variables with respect to the Calogero-Moser flow
but not with respect to the Gaudin flows.

Let us compute analogues of the Gaudin Hamiltonians from (\ref{w62})
in the Calogero model. The Hamiltonian $H_0$ (\ref{w63})
 \beq\label{w66}
 \begin{array}{c}
  \displaystyle{
H_0=\frac12\,\tr(\Lambda^2)=\frac12\,\tr({\ti Q}^2)=\frac12\,\tr(\Psi{\ti Q}\Psi^{-1})=
\frac12\,(\tr({L}^{\hbox{\tiny{CM}}})^2)=
\frac12\sum\limits_{i=1}^N p_i^2-\sum\limits_{i<j}^N\frac{\nu^2}{(q_i-q_j)^2}
   }
 \end{array}
 \eq
is the Calogero-Moser Hamiltonian. The Hamiltonians $H_a$ (\ref{w58}) computed through $L(z)$ (\ref{rs301})
vanish:
 \beq\label{w67}
 \begin{array}{c}
  \displaystyle{
H^{\hbox{\tiny{G}}}_a=\nu\tr(L^{\hbox{\tiny{CM}}} \bar\mO^a)+
\nu^2\sum\limits_{c:c\neq a}^N\frac{\tr(\bar\mO^a\bar\mO^c)}{q_a-q_c}=0
   }
 \end{array}
 \eq
 since from (\ref{rs2}) we have
 \beq\label{w671}
 \begin{array}{c}
  \displaystyle{
 \tr(\bar\mO^a\bar\mO^c)=1-\delta_{ac}
   }
 \end{array}
 \eq
 and
 \beq\label{w672}
 \begin{array}{c}
  \displaystyle{
 \tr(L^{\hbox{\tiny{CM}}} \bar\mO^a)=-\sum\limits_{c:c\neq a}^N\frac{\nu}{q_a-q_c}\,.
   }
 \end{array}
 \eq
 Summarizing, we see that using the described above approach,
  the last line of (\ref{w04}) has no direct analogue at classical level ($H^{\hbox{\tiny{G}}}_a=0$ here).

 Let us proceed to the fictitious Schlesinger systems. The Lax equation
 ${\dot L}^{\hbox{\tiny{CM}}}=[L^{\hbox{\tiny{CM}}},M^{\hbox{\tiny{CM}}}]$
can be equivalently written in the form
 \beq\label{w68}
 \begin{array}{c}
  \displaystyle{
 \p_t L^{\hbox{\tiny{CM}}} -\p_z M^{\hbox{\tiny{CM}}}=[L^{\hbox{\tiny{CM}}},M^{\hbox{\tiny{CM}}}]
   }
 \end{array}
 \eq
 or
 \beq\label{w69}
 \begin{array}{c}
  \displaystyle{
[\p_z+L^{\hbox{\tiny{CM}}},\p_t+M^{\hbox{\tiny{CM}}}]=0
   }
 \end{array}
 \eq
since $M^{\hbox{\tiny{CM}}}$ is independent of $z$. Then we deal with the connection $\p_z+L^{\hbox{\tiny{CM}}}$
instead of the Lax matrix $L^{\hbox{\tiny{CM}}}$. Perform the gauge transformations (\ref{rs1})-(\ref{rs301})
with this connection. First,
 \beq\label{w691}
 \begin{array}{c}
  \displaystyle{
\p_z+L^{\hbox{\tiny{CM}}}\rightarrow \p_z+L'(z)= (z-Q)(\p_z+L^{\hbox{\tiny{CM}}})(z-Q)^{-1}\,,
   }
 \end{array}
 \eq
 so that
 \beq\label{w692}
\displaystyle{
	L'(z)
	=L^{\hbox{\tiny{CM}}} + \sum\limits_{a = 1}^{N}\frac{{\mO}^{a}}{z-q_{a}}\,,\quad \mO^a\in\Mat\,,
}
\eq
where
 \beq\label{w693}
\displaystyle{
\mO^{a}_{ij} = -\nu(1-\delta_{ij})\delta_{aj}-\delta_{ij}\delta_{aj}\,,\quad
a,i,j=1,...,N\,.
}
\eq
Then similarly to (\ref{rs3}) we define
 \beq\label{w6921}
\displaystyle{
	L(z)=\Psi^{-1}L'(z)\Psi
	={\ti Q} + \sum\limits_{a = 1}^{N}\frac{{\Psi^{-1}\mO}^{a}\Psi}{z-q_{a}}\,,\quad \mO^a\in\Mat\,,
}
\eq
 The calculation of the Gaudin-Schlesinger Hamiltonians through $L(z)$ (\ref{w694}) in this case is similar to (\ref{w67})-(\ref{w672})
 but $\nu\bar\mO^a$ should be replaced with $\mO^a$. Since $\tr(\mO^a\mO^b)=\nu^2(1-\delta_{ab})+\delta_{ab}$
 and
 \beq\label{w694}
 \begin{array}{c}
  \displaystyle{
 \tr(L^{\hbox{\tiny{CM}}} \mO^a)=-\nu p_a-\sum\limits_{c:c\neq a}^N\frac{\nu^2}{q_a-q_c}
   }
 \end{array}
 \eq
 we have
 \beq\label{w70}
 \begin{array}{c}
  \displaystyle{
H^{\hbox{\tiny{Sch}}}_a=\nu\tr(L^{\hbox{\tiny{CM}}} \mO^a)+
\sum\limits_{c:c\neq a}^N\frac{\tr(\mO^a\mO^c)}{q_a-q_c}=-\nu p_a\,.
   }
 \end{array}
 \eq
 The latter relation is similar to the last line of (\ref{w04}).
Together with (\ref{w64})-(\ref{w65}) we call it the classical-classical counterpart of
quantum-classical duality (\ref{w04}).

Finally, let us mention that in the non-autonomous case the symplectic structure acquires
an additional term $\Delta\omega=\sum\limits_{a=1}^N dH^{\hbox{\tiny{Sch}}}_a\wedge dz_a$,
where $H^{\hbox{\tiny{Sch}}}_a$ are the Gaudin-Schlesinger Hamiltonians.
In our case $\Delta\omega=-\nu\sum\limits_{a=1}^N dp_a\wedge dq_a$, that is
the classical-classical duality assumes the Gaudin-Schlesinger Hamiltonians
to be canonically conjugated to moving poles in a natural way.

       %

\section{Appendix A: trigonometric $r$-matrix}\label{secA}
\def\theequation{A.\arabic{equation}}
\setcounter{equation}{0}

Trigonometric Gaudin model is described by the trigonometric classical $r$-matrix \cite{BD}:
  \beq\label{w351}
  \begin{array}{c}
    \displaystyle{
 r^{\hbox{\tiny{xxz}}}_{12}(\zeta)=\coth(\zeta)\sum\limits_{i=1}^N E_{ii}\otimes E_{ii}+
 \frac{1}{\sinh(\zeta)}\sum\limits_{i<j}^N
 \Big( E_{ij}\otimes E_{ji}e^\zeta + E_{ji}\otimes E_{ij}e^{-\zeta}\Big)\in\Mat^{\otimes 2}\,,
 }
 \end{array}
 \eq
which is the classical
limit of the quantum XXZ $U_q({\rm gl}_N)$ $R$-matrix. Here $\{E_{ij},\ i,j=1,...,N\}$ is the standard
matrix basis in $\Mat$. This $r$-matrix is skew-symmetric $r_{12}(-\zeta)=-r_{21}(\zeta)$ and satisfies the classical Yang-Baxter equation
  \beq\label{w352}
  \begin{array}{c}
    \displaystyle{
 [r_{12}(\zeta_1-\zeta_2),r_{23}(\zeta_2-\zeta_3)]+
  [r_{12}(\zeta_1-\zeta_2),r_{13}(\zeta_1-\zeta_3)]+
   [r_{13}(\zeta_1-\zeta_3),r_{23}(\zeta_2-\zeta_3)]=0\,.
 }
 \end{array}
 \eq
The Lax matrix is defined as
  \beq\label{w353}
  \begin{array}{c}
    \displaystyle{
 L^{\hbox{\tiny{xxzG}}}(\zeta)=\Lambda+\sum\limits_{k=1}^M \tr_2\Big(r^{\hbox{\tiny{xxz}}}_{12}(\zeta-\zeta_k)S_2^k\Big)\,,\quad S^k_2=1_N\otimes S^k\,,
 \quad S^k\in\Mat\,,\ k=1,...,M\,,
 }
 \end{array}
 \eq
where $\Lambda={\rm diag}(\la_1,...,\la_N)$ is a diagonal twist matrix, $\tr_2$ is a trace over the second tensor component, so that
  \beq\label{w354}
  \begin{array}{c}
    \displaystyle{
\tr_2\Big(r^{\hbox{\tiny{xxz}}}_{12}(\zeta)S_2\Big)=\coth(\zeta)\sum\limits_{i=1}^N E_{ii} S_{ii}+
\frac{e^\zeta}{\sinh(\zeta)}\sum\limits_{i<j}^N E_{ij} S_{ij}+
\frac{e^{-\zeta}}{\sinh(\zeta)}\sum\limits_{i>j}^N E_{ij} S_{ij}\,.
 }
 \end{array}
 \eq
Positions of simple poles $\zeta_k$, $k=1,...,M$ are called the marked points or punctures on the corresponding
complex curve with coordinate $\zeta$.

Let us transform the $r$-matrix (\ref{w351}) to the form which is convenient for the duality. First,
rewrite (\ref{w351}) in multiplicative variable $z=e^{2\zeta}$:
  \beq\label{w355}
  \begin{array}{c}
    \displaystyle{
 r^{\hbox{\tiny{xxz}}}_{12}(z)=
 \frac{z+1}{z-1}\sum\limits_{i=1}^N E_{ii}\otimes E_{ii}
 +\sum\limits_{i\neq j}^N E_{ij}\otimes E_{ji} \Big(\delta_{i<j}\frac{2z}{z-1}+\delta_{i>j}\frac{2}{z-1}\Big)=
  }
  \\
    \displaystyle{
 =\sum\limits_{i,j=1}^N
 E_{ij}\otimes E_{ji}\Big(\frac{z+1}{z-1}+{\rm sign}(j-i) \Big)\,,
 \qquad
 {\rm sign}(k)=(1-\delta_{k0})(\delta_{k>0}-\delta_{k<0})\,.
 }
 \end{array}
 \eq
Then
  \beq\label{w356}
  \begin{array}{c}
    \displaystyle{
 -\frac{1}{2}\,r^{\hbox{\tiny{xxz}}}_{12}(z^{-1})=
 \frac12\frac{z+1}{z-1}\sum\limits_{i=1}^N E_{ii}\otimes E_{ii}
 +\sum\limits_{i\neq j}^N E_{ij}\otimes E_{ji} \Big(\delta_{i<j}\frac{1}{z-1}+\delta_{i>j}\frac{z}{z-1}\Big)\,.
 }
 \end{array}
 \eq
Finally, define\footnote{In (\ref{w357}) we added a term proportional to $\sum\limits_{i=1}^N E_{ii}\otimes E_{ii}$. This
is the so-called twist transformation of the classical $r$-matrix. See e.g. Lemma 1 in Section 4 of \cite{LOSZ}.}
  \beq\label{w357}
  \begin{array}{c}
    \displaystyle{
 r_{12}(z)=-\frac{1}{2}\,r^{\hbox{\tiny{xxz}}}_{12}(z^{-1})-\frac12\sum\limits_{i=1}^N E_{ii}\otimes E_{ii}
 =\sum\limits_{i,j=1}^N E_{ij}\otimes E_{ji} \Big(\frac{1}{z-1}+\delta_{i>j}\Big)\,.
 }
 \end{array}
 \eq
This is the $r$-matrix (\ref{w3571}).
Notice that this $r$-matrix is not skew-symmetric (i.e. $r_{12}(z^{-1})\neq -r_{21}(z)$). It satisfies the classical
Yang-Baxter equation for non skew-symmetric $r$-matrices:
  \beq\label{w358}
  \begin{array}{c}
    \displaystyle{
 [r_{12}(z_1/z_2),r_{23}(z_2/z_3)]+
  [r_{12}(z_1/z_2),r_{13}(z_1/z_3)]+
   [r_{32}(z_3/z_2),r_{13}(z_1/z_3)]=0\,.
 }
 \end{array}
 \eq
%

\section{Appendix B: gauge transformation for XXZ Gaudin model}\label{secB}
\def\theequation{B.\arabic{equation}}
\setcounter{equation}{0}

Here we prove (\ref{w3611})-(\ref{w362}).

\noindent{\bf Lemma}
{\em
Let ${\bar \mS}$ be a matrix with elements ${\bar \mS}=\delta_{i>j}\mS_{ij}$, where
\begin{equation}\label{p1}
	S_{ij} = \sum\limits_{k = 1}^M\xi_{i}^{k}\eta_{j}^{k}\,.
\end{equation}	
Then
\begin{equation}\label{p2}
	g^{-1}(\Lambda+{\bar \mS})g=\Lambda\,
\end{equation}
where the $g\in\Mat$ with elements
\begin{equation}\label{p3}
		g_{ij}=\delta_{ij}+\frac{\delta_{i>j}}{\la_j-\la_i}
	\xi_i
	\Big( 1_M+\frac{ \eta_{i-1}\otimes\xi_{i-1} }{\la_j-\la_{i-1}} \Big)
	\dots
	\Big( 1_M+\frac{ \eta_{j+1}\otimes\xi_{j+1} }{\la_j-\la_{j+1}} \Big)
	\eta_j\,.
\end{equation}
}
\begin{proof}
Rewrite \eqref{p2} as
\begin{equation}\label{p4}
		{\bar \mS}g = \left[g,\Lambda\right]\,.
\end{equation}
Define $g_{ij} = \delta_{ij}+\delta_{i > j}G_{ij}$, and let us treat \eqref{p4} as an equation for $G$.
This leads to the recursive relations:
\begin{equation}\label{p5}
	G_{j+1,j} = \frac{\bar \mS_{j+1,j}}{\lambda_{j}-\lambda_{j+1}}\,,
\end{equation}	
\begin{equation}\label{p6}
	G_{j+k,j}\left(\lambda_{j}-\lambda_{j+k}\right) = \bar \mS_{j+k,j} + \sum\limits_{d = 1}^{k-1}{\bar\mS}_{j+k,j+d}G_{j+d,j}\,.
\end{equation}
We need to prove that
\begin{equation}\label{p7}
	G_{j+k,j} = \frac{1}{\lambda_{j}-\lambda_{j+k}}\xi_{j+k}\prod\limits_{p = j+1}^{j+k-1}\left(1_{M}+\frac{\eta_{p}\otimes\xi_{p}}{\lambda_{j}-\lambda_{p}}\right)\eta_{j}
\end{equation}
is the solution of equations (\ref{p5})-(\ref{p6}).
Substituting (\ref{p6}) in (\ref{p7}), one obtains
\begin{equation}\label{dp1}
\xi_{j+k}\prod\limits_{p = j+1}^{j+k-1}\left(1_{M}+\frac{\eta_{p}\otimes\xi_{p}}{\lambda_{j}-\lambda_{p}}\right)\eta_{j} = \xi_{j+k}\left(1_{M} + \sum\limits_{d = 1}^{k-1}\frac{\eta_{j+d}\xi_{j+d}}{\lambda_{j}-\lambda_{j+d}}\prod\limits_{p = j+1}^{j+d-1}\left(1_{M}+\frac{\eta_{p}\otimes\xi_{p}}{\lambda_{j}-\lambda_{p}}\right) \right)\eta_{j}.
\end{equation}
It is true for $k = 1$ and it is sufficient to prove by induction that
\begin{equation}\label{p8}
	1_{M} + \sum\limits_{d = 1}^{k-1}\frac{\eta_{j+d}\xi_{j+d}}{\lambda_{j}-\lambda_{j+d}}\prod\limits_{p = j+1}^{j+d-1}\left(1_{M}+\frac{\eta_{p}\otimes\xi_{p}}{\lambda_{j}-\lambda_{p}}\right) = \prod\limits_{p = j+1}^{i+k-1}\left(1_{M}+\frac{\eta_{p}\otimes\xi_{p}}{\lambda_{j}-\lambda_{p}}\right).
\end{equation}
Assume that \eqref{p8} holds for $k = t$, then for  $k = t+1$
\begin{align}\label{p9}
	&1_{M} + \sum\limits_{d = 1}^{t}\frac{\eta_{j+d}\xi_{j+d}}{\lambda_{j}-\lambda_{j+d}}\prod\limits_{p = j+1}^{j+d-1}\left(1_{M}+\frac{\eta_{p}\otimes\xi_{p}}{\lambda_{j}-\lambda_{p}}\right) = \prod\limits_{p = j+1}^{j+t-1}\left(1_{M}+\frac{\eta_{p}\otimes\xi_{p}}{\lambda_{j}-\lambda_{p}}\right) + \notag \\&+
	\frac{\eta_{j+t}\xi_{j+t}}{\lambda_{j}-\lambda_{j+t}}\prod\limits_{p = j+1}^{j+t-1}\left(1_{M}+\frac{\eta_{p}\otimes\xi_{p}}{\lambda_{j}-\lambda_{p}}\right) = \prod\limits_{p = j+1}^{j+t}\left(1_{M}+\frac{\eta_{p}\otimes\xi_{p}}{\lambda_{j}-\lambda_{p}}\right).
\end{align}
This proves that \eqref{p7} is a solution of equations (\ref{p5})-(\ref{p6}).
\end{proof}

\subsection*{Acknowledgments}



This work was performed at the Steklov International Mathematical Center and supported by the Ministry of Science and Higher Education of the Russian Federation (agreement no. 075-15-2022-265). The work of Rostislav Potapov was supported by Basis foundation.


\begin{small}

\end{small}

\end{document}